\documentstyle[aps,prd,epsfig]{revtex}
\newcommand{\pbp}{\langle\bar\psi\psi\rangle}

\preprint{          Preprint numbers: \ \
          UUHEP 99/2 \ \
          NORDITA-99/25HE \ \
}
\begin{document}
\draft

\title{Critical Behavior in $N_t = 4$ Staggered Fermion Thermodynamics}

\author{Claude Bernard}
\address{
Department of Physics, Washington University, 
St.~Louis, MO 63130, USA
}
\author{Carleton DeTar}
\address{
Department of Physics, University of Utah, 
Salt Lake City, UT 84112, USA
}
\author{Steven Gottlieb}
\address{
Indiana University, Bloomington, IN 47405, USA
}
\author{Urs M.~Heller}
\address{
SCRI, The Florida State University, Tallahassee, FL 32306-4130, USA
}
\author{James Hetrick} 
\address{
University of the Pacific, Stockton, CA 95211, USA
}
\author{Kari Rummukainen}
\address{
NORDITA, Blegdamsvej 17, DK-2100 Copenhagen \O, Denmark
}

\author{Robert L.~Sugar}
\address{
Department of Physics, University of California, 
Santa Barbara, CA 93106, USA
}
\author{Doug Toussaint}
\address{
Department of Physics, University of Arizona, Tucson, AZ 85721, USA
}
\date{\today}
\maketitle
\begin{abstract}
  Quantum chromodynamics with two zero mass flavors is expected to
  exhibit a phase transition with O(4) critical behavior.  Fixing the
  universality class is important for phenomenology and for
  facilitating the extrapolation of simulation data to physical quark
  mass values.  Other groups have reported results
  from lattice QCD simulations with dynamical staggered quarks at $N_t
  = 4$, which suggest a departure from the expected critical behavior.
  We have pushed simulations to the largest volumes and smallest quark
  mass to date.  Strong discrepancies in critical exponents and the scaling
  equation of state persist.
\end{abstract}
\pacs{}

\section{Introduction} 

It is generally expected that two-flavor QCD undergoes a high
temperature chiral-symmetry-restoring phase transition at zero quark
mass with O(4) critical behavior\cite{PW}.  Should the axial anomaly
disappear simultaneously with the phase transition, the Pisarski
Wilczek scenario then suggests a fluctuation-driven first order phase
transition.  Verifying these expectations is important for understanding
the phenomenology of the transition and for facilitating an
extrapolation of simulation data to physical quark masses.  Since the
staggered fermion scheme breaks the anomaly explicitly at nonzero
lattice spacing, lattice QCD at fixed $N_t$ with staggered fermions,
as a statistical system in its own right, is similarly expected to
exhibit at least O(2) universality, with O(4) or a fluctuation-driven
first-order phase transition emerging in the continuum limit $N_t
\rightarrow \infty$.  At $N_t = 4$ the lattice spacing is coarse
enough that, if there is a critical point at zero quark mass, O(2) is
the only likely option.

The standard test of universality compares critical exponents.
Comparing the critical scaling function itself gives further insight.
To test for the expected universality we use the standard
correspondence between QCD variables and O(N) spin variables, which
identifies quark mass $m_q/T$ with magnetic field $h$, inverse gauge
coupling $6/g^2$ with temperature $T/T_c(0)$, chiral condensate $\pbp$
with magnetization $M$, and the action (plaquette) with the energy
density. A critical point is expected to occur at zero quark mass and
nonzero coupling $6/g^2(0)$.  For studies at fixed $N_t$, therefore,
we define\cite{KL}
\begin{eqnarray}
  h &=& am_q N_t \cr
  t &=& 6/g^2 - 6/g_c^2 \propto T/T_c - 1 
\end{eqnarray}

Critical scaling theory predicts that for small quark masses
we have the Fisher scaling relation \cite{Amit}
\begin{equation}
  \pbp h^{-1/\delta} = f_{\rm QCD}(x) = c_y
  f_G(c_x x) 
  \label{eq:scaling_fcn}
\end{equation}
where $x = t h^{-1/\beta\delta}$, $c_x$ and $c_y$ are scale constants,
$f_{QCD}(x)$ is the critical scaling function for QCD and $f_G(x)$ is
that for the appropriate universality class $G$.  Only the scale
constants $c_x$ and $c_y$ are adjustable.  Some critical exponents are
given in Table \ref{tab:crit_exp}.  Outside the Ginzburg scaling
region, by definition, there are appreciable nonleading, nonscaling
contributions to $\pbp$, analytic in $t$ and $h$.  In addition to
corrections analytic in $t$ and $h$, there are correction terms with
subleading exponents, universal and nonanalytic in $t$ and $h$.  The
mean-field scaling function is known exactly.  For the O(4) scaling
function we use results of a numerical simulation\cite{O4}.

There is a similar scaling relation for the energy density.  In QCD
the energy density (plaquette) is dominated by gluon degrees of
freedom, which are indirectly affected by the chiral singularity. So
apparently there is a much larger analytic contribution.
Consequently, we have found the plaquette observable much less useful
for testing critical scaling.  Here we concentrate on the scaling of
$\pbp$.

In the next section we discuss an analysis of finite size effects,
present a determination of some critical exponents, and compare our
results with the critical scaling function.  In the concluding section
we suggest reasons for the discrepancies observed.  A preliminary
version of this study was presented at Lattice '97 \cite{CD97}.

\section{Analysis of critical behavior}

Our data set extends an old sample on lattice sizes $L^3\times N_t$
with $N_t = 4$, $L = 8$ and quark masses $am_q = 0.025$ and 0.0125,
which was generated with the standard one-plaquette gauge action plus
two-flavor staggered fermion action.  Our new simulations decrease the
quark mass to $am_q = 0.008$ and increase the spatial lattice size $L$
to 24 (aspect ratio 6). We also reanalyzed old data at $N_t = 6$, 8,
and 12 \cite{Nt12}.  The old data, unfortunately, are limited to
aspect ratio $L/N_t = 2$.  The extent of our $N_t = 4$ data sample is
given in Table \ref{tab:data_set}.  Included in this table are values
for global observables.  For equilibration we typically dropped the
first 300 molecular dynamics time units of each run.

Over the range of nonzero quark masses considered, there appears to be
no phase transition -- only a crossover, as illustrated for $am_q =
0.008$ in Fig.~\ref{fig:pbp_m008}.  Evidently, however, the crossover
steepens as the lattice volume is increased.  The crossover, or
``pseudo-critical point'' is signaled by a peak in a susceptibility
for any lattice size.  For example the mixed plaquette/chiral
condensate susceptibility, corresponding to the slope in
Fig.~\ref{fig:pbp_m008},
\begin{equation}
\chi_{mt} = \frac{\partial \pbp}{\partial (6/g^2)}
\end{equation} 
is plotted in Fig.~\ref{fig:pbpA} for the $16^3\times 4$ lattice.
Here, as well as in Fig.~\ref{fig:pbp_m008}, we use multihistogram
reweighting to interpolate the data from the simulation points and
locate the peak.  The error analysis was performed with the jackknife
method, which enables us to obtain reliable error estimates for both
the peak height and location.  

The peak location (crossover coupling $6/g_{\rm pc}^2$) shows little
variation in lattice size for $L > 8$.  For example the peak location
in $\chi_{mt}$ shifts from 5.2605(10) to 5.2623(6) as $L$ increases
from 12 to 24, a scarcely significant change.  It also shows little
variation among the susceptibilities chosen.  For example the peak
location varies by $\pm 0.001$ over the susceptibilities considered.
In all cases we take the result from the largest volume and assign an
error of 0.002.  Close to the critical point the peak position occurs
at a fixed value of the scaling variable $x = x_{\rm pc}$, so we have
the scaling relation \cite{KL}
\begin{equation}
  6/g_{\rm pc}^2 = 6/g_{\rm pc}(0)^2 + x_{\rm pc} (am_qN_t)^{1/\delta\beta}.
\end{equation}
Shown in Fig.~\ref{fig:pseudo_crit_scaling} is the trajectory of the
pseudocritical point, fitted to both $O(4)$ and mean-field
predictions.  Both fits are good.  An $O(2)$ fit would do equally
well.  Such agreement was first found by Karsch and Laermann and
inspired hope that the simulations had entered the scaling region
\cite{KL}.

Problems with scaling were uncovered in studies at larger volumes
\cite{U96,B96,JLQCD97}.  In Figs.~\ref{fig:pbpA_ext_0125} and
\ref{fig:pbpA_ext_008} we plot the peak height of the $\chi_{mt}$
susceptibility for two fixed quark masses.  The increase in peak
height with increasing volume reflects the steepening trend seen, for
example, in Fig.~\ref{fig:pbp_m008}. It is necessary to extrapolate to
infinite volume at each quark mass before checking scaling.  We start
by assuming the conventional scenario, in which the critical point
occurs at $am_q = 0$.  Then at nonzero mass, the susceptibility has a
finite limit at large volume.  We make an {\it ad hoc} choice for an
extrapolation formula with the result shown in
Figs.~\ref{fig:pbpA_ext_0125} and \ref{fig:pbpA_ext_008}.
\begin{equation}
   \chi^{\rm max}_{mt}(L) = \chi^{\rm max}_{mt}(\infty) + b/L.
\end{equation}
%
The $am = 0.0125$ data covers the largest range of lattice sizes.
Varying the inverse power of $L$ from 1/2 to 1 to 2 gives a slight
preference for $1/L$ at this mass.  Given the uncertainties in the
values themselves, we feel it is safe to use any of these
extrapolations, and we have chosen $1/L$ for all masses.

The extrapolated peak height of the $\chi_{mt}$ susceptibility is
expected to scale with decreasing quark mass.  For this
susceptibility, the expected scaling relation is
\begin{equation}
  \chi^{\rm max}_{mt} \sim (am_q)^{(\beta - 1)/\beta\delta}.
\end{equation}
%
We compare this prediction with results from our analysis in
Fig.~\ref{fig:pbpA_ext}.  Also shown are similarly extrapolated JLQCD
values \cite{JLQCD97}.  If we include all points in the fit, the
scaling exponent is $-1.08(8)$, compared with an $O(4)$ prediction of
$-0.33$ --- a clear disagreement, corroborating results of the JLQCD
and Bielefeld groups \cite{U96,B96,JLQCD97}.  However, it is evident
in Fig.~\ref{fig:pbpA_ext} that at the three lightest masses this
observable alone does not exclude O(4).  To test sensitivity to our
extrapolation formula, we carried out the same analysis, replacing
$1/L$ by $1/\sqrt{L}$ and $1/L^2$.  The resulting scaling exponents
are $-1.24(11)$ and $-0.94(5)$, respectively, still clearly at
variance with $O(4)$ over the full mass range studied.

A similar fit of the plaquette susceptibility, $\chi_{tt} = \partial
\langle\Box\rangle/\partial(6/g^2)$, also including $1/L$-extrapolated
JLQCD results, yields a scaling exponent of $-0.78(7)$, while the
$O(4)$ prediction is $1/\delta - 2/(\delta\beta) + 1 = 0.13$.  The
$1/\sqrt{L}$ and $1/L^2$ extrapolations give $-0.95(10)$ and
$-0.64(5)$, respectively.

Because the crossover steepens so much with increasing lattice volume
and small quark mass, it is worthwhile looking for evidence for
two-phase metastability, signaling a first-order phase transition.
Figure \ref{fig:pbp.m008l24} shows the simulation time histories of
our runs at $am_q=0.008$.  While we certainly see long correlation
times, we see no evidence for a first order transition in these
histories.  In Fig.~\ref{fig:pbp_hot_cold} we show time histories from
hot and cold starts at $am_q=0.0125$ (two hot and two cold starts) and
$am_q=0.008$ at values of $6/g^2$ near the peaks of the
susceptibilities.  Again, there is no evidence for a first order
transition.

We turn next to an analysis of the critical scaling function, given by
Eq (\ref{eq:scaling_fcn}).  Here, again, we assume that we are in the
scaling region.  The analysis then depends on which critical exponents
we adopt.  Using $O(4)$ critical exponents from Kanaya and Kaya
\cite{KK}, we construct $f_{\rm QCD}(x)$ and compare with the scaling
function $f_{O(4)}(x)$ for O(4) \cite{O4} in
Fig.~\ref{fig:pbpnt4.change.5225}.  Essentially all of the data lie at
positive $x$, which permits a log-log plot.  Vertical and horizontal
displacements of the log-log scaling curves correspond to adjusting
$c_x$ and $c_y$.  Clearly, no such displacement would result in good
agreement.  The newer data are plotted with octagons and squares.  We
observe: (1) The QCD curve falls with increasing steepness as the
quark mass is decreased.  Since the slope of the curve at the
crossover gives the peak height of the $\chi_{mt}$ susceptibility, the
disagreement there is consistent with the observed lack of scaling of
the peak height itself. (2) The new data at larger volume and smaller
quark mass show generally worse agreement with the $O(4)$ scaling
curve.  (3) The crossover regions, indicated in the QCD results by
line segments and in the $O(4)$ scaling function by a dashed line, are
far from agreeing.

We show a similar comparison of the QCD scaling function with the
mean-field prediction in Fig.~\ref{fig:pbpnt4.change.5235.mf}.  Again
the disagreement is significant.  Although we have not measured the
$O(2)$ scaling function, so cannot make a direct comparison, given the
close similarity of the critical exponents with $O(4)$, we do not
expect any improvement with that choice.

We conclude that if the $N_t = 4$ theory falls in the $O(2)$ or $O(4)$
universality class, simulations at present masses do not reach the
critical scaling region.  Furthermore, as the quark mass is decreased
over the present range, disagreement with scaling predictions worsens,
offering little hope that we might be getting closer.

A similar analysis at larger $N_t$ is shown in Fig.~\ref{fig:pbp_o4}.
Perhaps there is improvement with increasing $N_t$.  However, our $N_t
= 12$ sample includes data only at a single quark mass, making it the
weakest test.  Furthermore, for $N_t > 4$ we have no results for $L >
2 N_t$, where we first encountered difficulties at $N_t = 4$.

\section{Discussion and Speculations}

We have seen that new simulations at smaller quark mass and larger
volume at $N_t = 4$ have raised doubts about the extent of the
previously observed agreement between QCD and $O(4)$ \cite{KL,Nt12}.
The conventional staggered fermion action with the conventional choice
of scaling variables does not show good agreement with the $O(4)$ or
mean field scaling functions at present quark masses and
temperatures. (Wilson quarks with an improved gauge action seem to
behave very differently \cite{Wscale}.)

Believers in the conventional sigma model scenario could argue that
the critical region is attained only when $\pi$ and $\sigma$
correlation lengths are considerably greater than $1/T_c$.  Only in
that case is the reduction of QCD to a three-dimensional sigma model
well justified.  Here, typically, these correlation lengths are
smaller than $1/T_c$.  Still, the observed worsening of the agreement
with decreasing quark mass is disturbing.

Recent results from simulations of the conventional $N_t = 4$
staggered fermion action, augmented by a four-fermi term (``chiral
QCD'') permit another speculation\cite{KLS4}.  With the additional
four-fermi interaction, Kogut, Laga\"e, and Sinclair are able to carry
out simulations at precisely zero quark mass.  They find evidence for
a first order phase transition at small four-fermi coupling.  A nearby
first order phase transition could spoil the approach to the critical
point.  Indeed, one cannot, therefore, rule out the possibility that
the first order phase transition extends to zero four-fermion coupling
for a small range of quark masses below the reach of our simulations.
In this case, one expects a critical end-point at a nonzero quark mass
$m_{qc}$ in the Ising or mean-field universality class.  At $N_t = 6$
the same group finds evidence for a crossover instead of a first-order
phase transition \cite{KLS6}.  Thus, one may speculate that the
conventional one-plaquette, staggered fermion action at $N_t = 4$ is
plagued by lattice artifacts large enough to obliterate the expected
$am_q = 0$ critical point, but these artifacts diminish at higher
$N_t$.

\acknowledgements

This work was supported by the U.S. Department of Energy under grants
DE--FG02--91ER--40661, 
DE--FG02--91ER--40628, 
DE--FG03--95ER--40894, 
DE--FG03--95ER--40906, 
DE--FG05--96ER--40979,  
DE--FG05--96ER--40979, 
and National Science Foundation grants
NSF--PHY96--01227 and     
NSF--PHY97--22022.        
Calculations were carried out through grants of computer
time from the NSF at NCSA and SDSC, and by the DOE at NERSC
and ORNL\@.  Some computations were carried out with the Indiana
University Paragon.

\begin{table}
\caption{Some critical exponents in three dimensions
\label{tab:crit_exp}
}
\begin{tabular}{ldddd}
  & $y_t$ & $y_h$ & $\delta$ & $\beta$ \\
\hline
MF     & 1.5    & 2.25   & 3     & 0.5 \\
$O(2)$ & 1.495  & 2.484  & 4.81  & 0.3455 \\
$O(4)$ & 1.337  & 2.487  & 4.851 & 0.3836 \\
$Z(2)$ & 1.61   & 2.5    & 5.0   & 0.31   
\end{tabular}
\end{table}
\begin{table}
\caption{Parameters in $N_t = 4$ data set and two global
observables. Run and step lengths are in molecular dynamics time
units.  \label{tab:data_set} }
\begin{tabular}{rdddrdd}
$L$ & $am_q$ & $6/g^2$ & step & length & plaquette & $\pbp$    \\
\hline
12 & 0.008  & 5.25    & .003 &  1965 & 1.455(2)    & 0.346(4)    \\
12 & 0.008  & 5.255   & .003 &  2200 & 1.467(4)    & 0.320(11)   \\
12 & 0.008  & 5.26    & .003 &  2130 & 1.506(6)    & 0.20(2)     \\
12 & 0.008  & 5.265   & .003 &  2170 & 1.512(4)    & 0.18(2)     \\
12 & 0.008  & 5.27    & .003 &  2045 & 1.532(2)    & 0.116(7)    \\
12 & 0.008  & 5.28    & .003 &  1965 & 1.5466(8)   & 0.075(2)    \\
12 & 0.0125 & 5.25    & .005 &  2920 & 1.4515(10)  & 0.364(2)    \\
12 & 0.0125 & 5.26    & .005 &  4630 & 1.467(2)    & 0.335(5)    \\
12 & 0.0125 & 5.27    & .005 &  7320 & 1.498(3)    & 0.254(11)   \\
12 & 0.0125 & 5.28    & .005 &  2820 & 1.5383(15)  & 0.129(5)    \\
12 & 0.025  & 5.27    & .01  &  2150 & 1.4652(9)   & 0.370(2)    \\
12 & 0.025  & 5.28    & .01  &  2075 & 1.483(2)    & 0.335(5)    \\
12 & 0.025  & 5.29    & .01  &  1975 & 1.512(4)    & 0.268(10)   \\
\hline
16 & 0.008  & 5.255   & .003 &  2460 & 1.4673(10)  & 0.321(3)    \\
16 & 0.008  & 5.26    & .003 &  1445 & 1.494(5)    & 0.24(2)	  \\
16 & 0.008  & 5.265   & .003 &  1825 & 1.520(3)    & 0.155(12)	  \\
16 & 0.008  & 5.27    & .003 &  1310 & 1.5346(8)   & 0.105(3)	  \\
16 & 0.0125 & 5.27    & .005 &  4700 & 1.500(4)    & 0.251(13)	  \\
16 & 0.0125 & 5.275   & .005 &  4900 & 1.530(2)    & 0.153(6)	  \\
\hline
24 & 0.008  & 5.255   & .003 &   950 & 1.4656(8)   & 0.326(2)    \\
24 & 0.008  & 5.26    & .003 &  1698 & 1.484(2)    & 0.276(5)    \\
24 & 0.008  & 5.263   & .003 &  1703 & 1.507(2)    & 0.202(8)	  \\
24 & 0.008  & 5.265   & .003 &  1702 & 1.5238(12)  & 0.140(5)	  \\
24 & 0.008  & 5.27    & .003 &  1700 & 1.5350(5)   & 0.104(2)	  \\
24 & 0.0125 & 5.265   & .005 &  1950 & 1.4747(6)   & 0.3208(15)  \\
24 & 0.0125 & 5.268   & .005 &  1760 & 1.487(2)    & 0.288(6)	  \\
24 & 0.0125 & 5.27    & .005 &  3126 & 1.502(2)    & 0.243(7)	  \\
24 & 0.0125 & 5.272   & .005 &  1760 & 1.5198(13)  & 0.186(5)	  \\
24 & 0.0125 & 5.275   & .005 &  1950 & 1.5295(11)  & 0.155(4)

\end{tabular}
\end{table}
%
\figure{
 \epsfig{bbllx=100,bblly=230,bburx=530,bbury=740,clip=,
         file=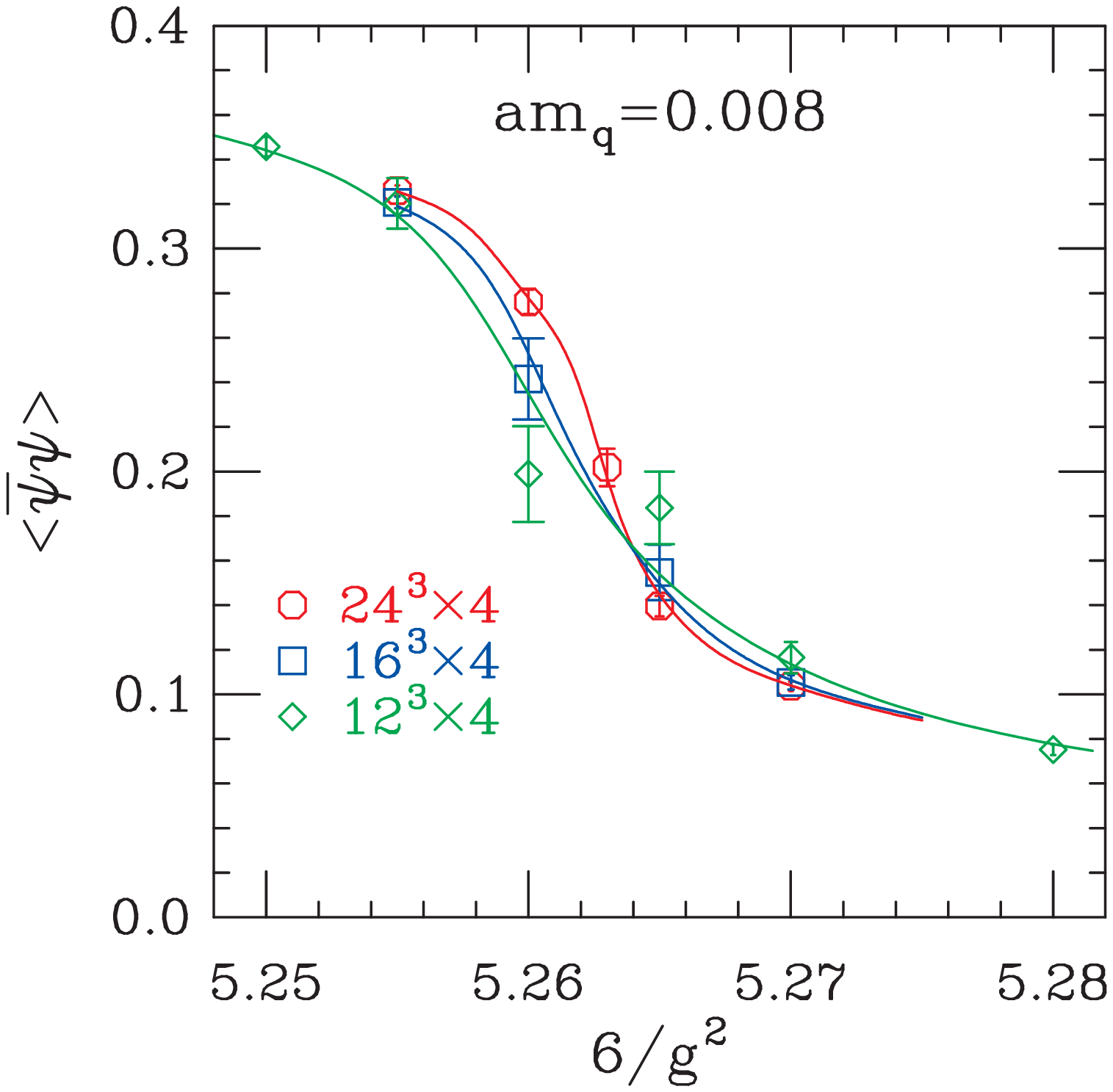,width=150mm}
\caption{Order parameter {\it vs} inverse gauge coupling for various
lattice sizes for $am_q = 0.008$.  Curves show results from
reweighting the data sample.
\label{fig:pbp_m008}
}
}
\figure{
 \epsfig{bbllx=100,bblly=230,bburx=530,bbury=740,clip=,
         file=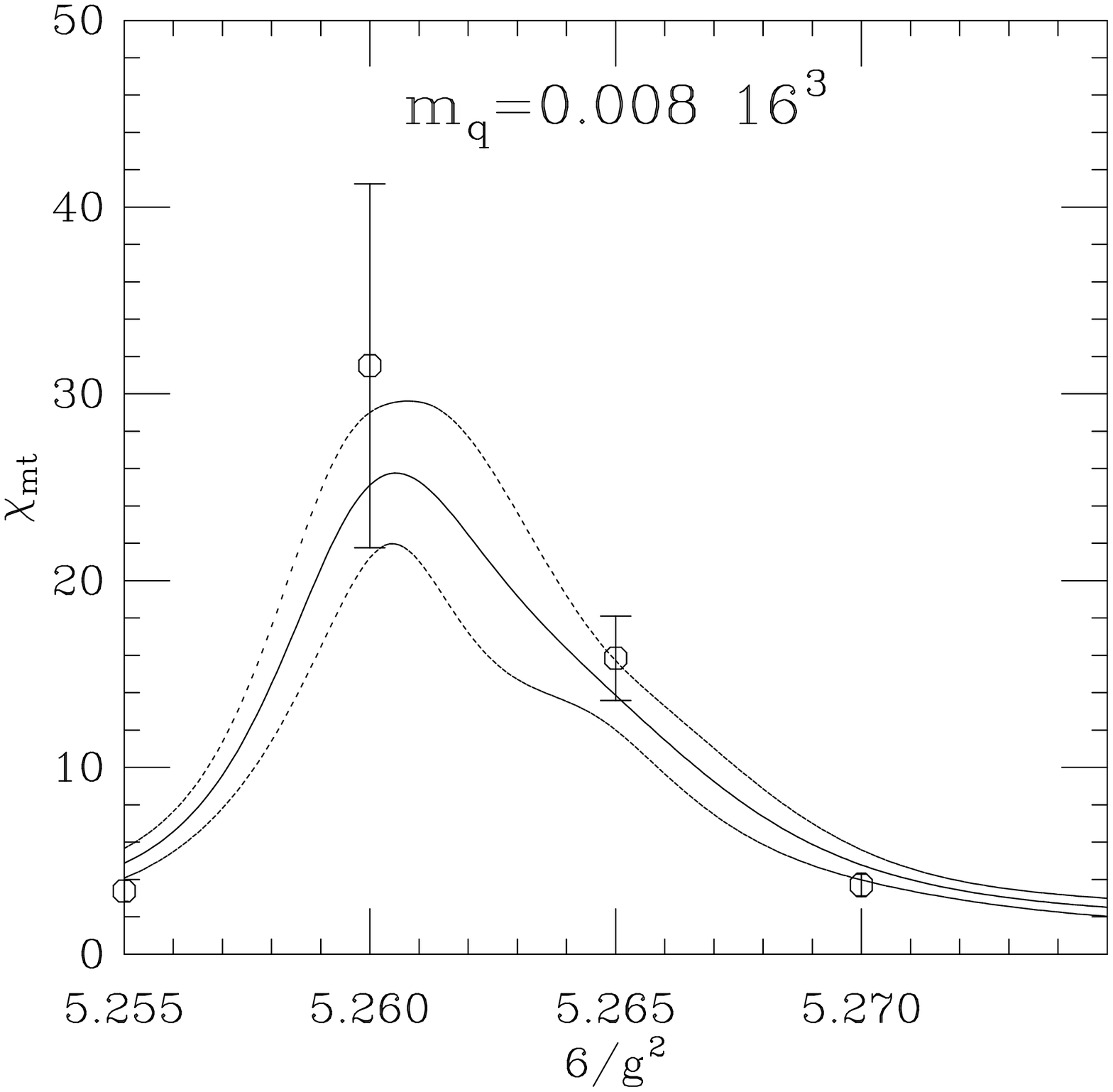,width=150mm}
\caption{The mixed $\pbp$--plaquette susceptibility $\chi_{mt}$ as a
function of inverse coupling $6/g^2$ for $am_q = 0.008$ on a $16^3\times
 4$ lattice.  Curves show results from reweighting together with
 one-standard-deviation bootstrap errors.
\label{fig:pbpA}
}
}
\figure{
 \epsfig{bbllx=100,bblly=230,bburx=530,bbury=740,clip=,
         file=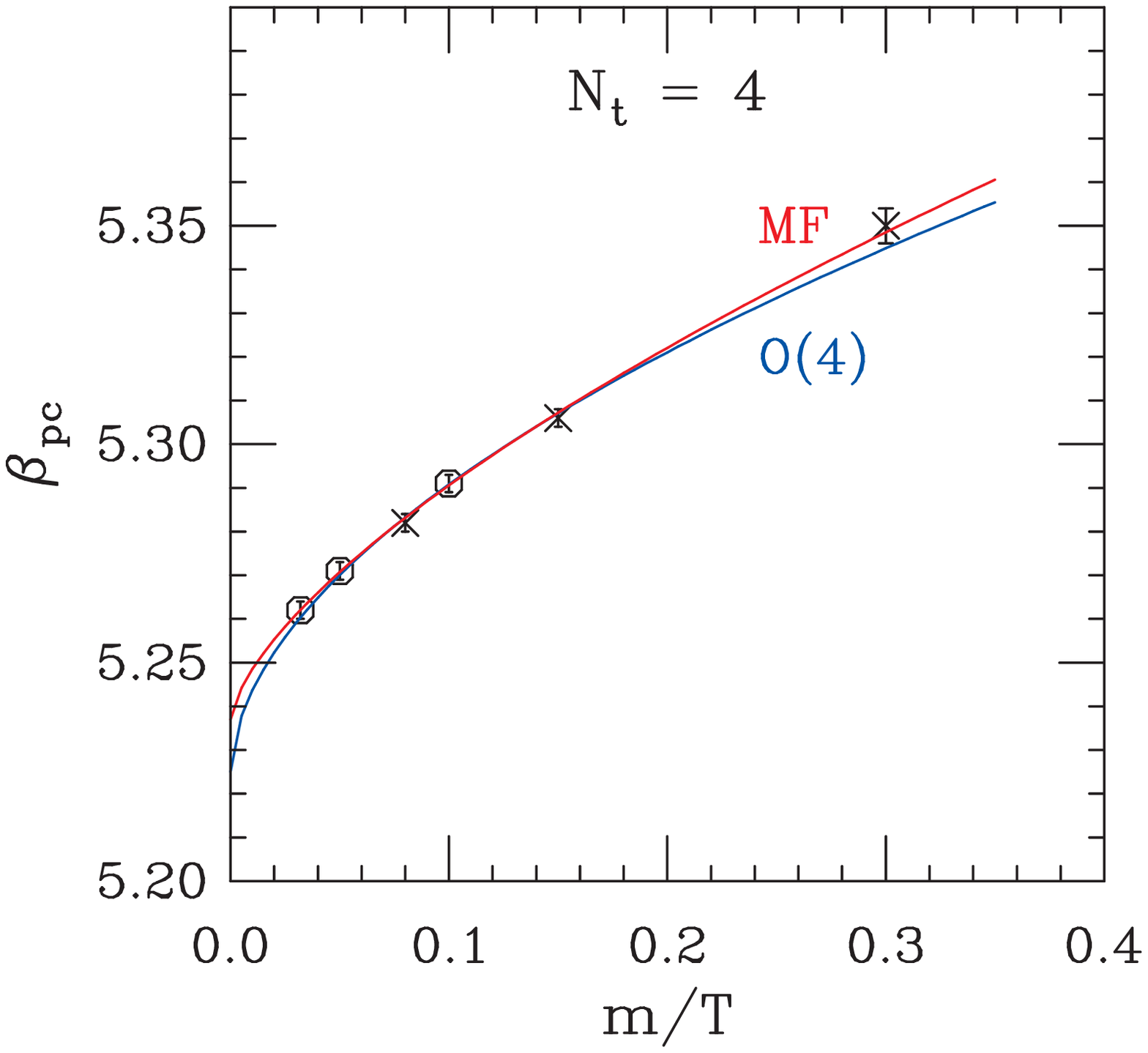,width=150mm}
\caption{Trajectory of the pseudocritical point $6/g^2_{\rm pc}$ as a
function of quark mass in units of temperature $m_q/T$ for $N_t = 4$.
Crosses indicate points from Karsch and Laermann \protect\cite{KL}.
Also shown are fits to both mean-field (upper curve) and $O(4)$ (lower
curve) scaling predictions.
\label{fig:pseudo_crit_scaling}
}
}
%
\figure{
 \epsfig{bbllx=100,bblly=230,bburx=530,bbury=740,clip=,
         file=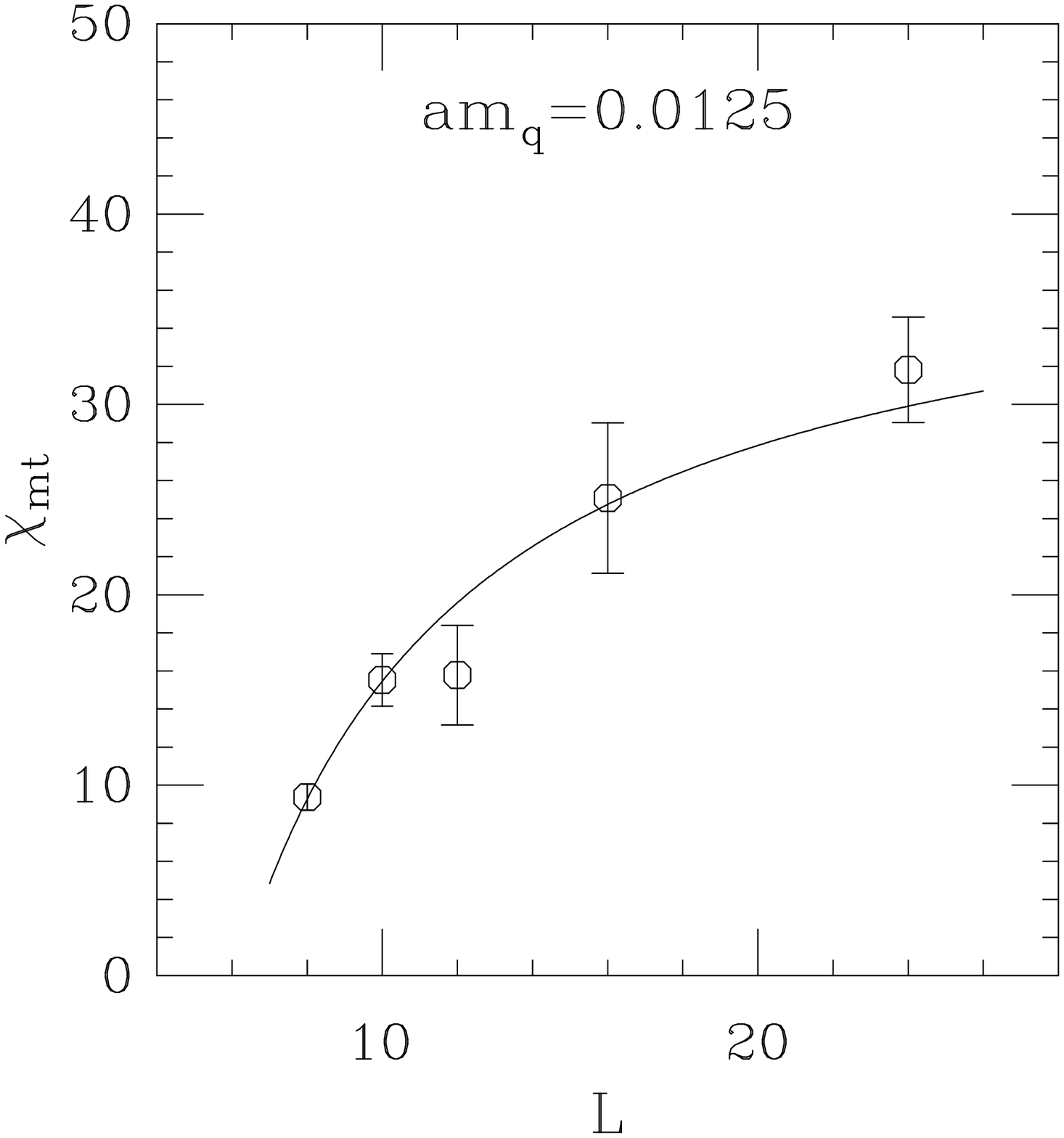,width=150mm}
\caption{Finite size analysis of the peak height in the mixed
$\pbp$--plaquette susceptibility $\chi_{mt}$ at $am_q = 0.0125$ for
$N_t = 4$.
\label{fig:pbpA_ext_0125}
}
}
%
\figure{
 \epsfig{bbllx=100,bblly=230,bburx=530,bbury=740,clip=,
         file=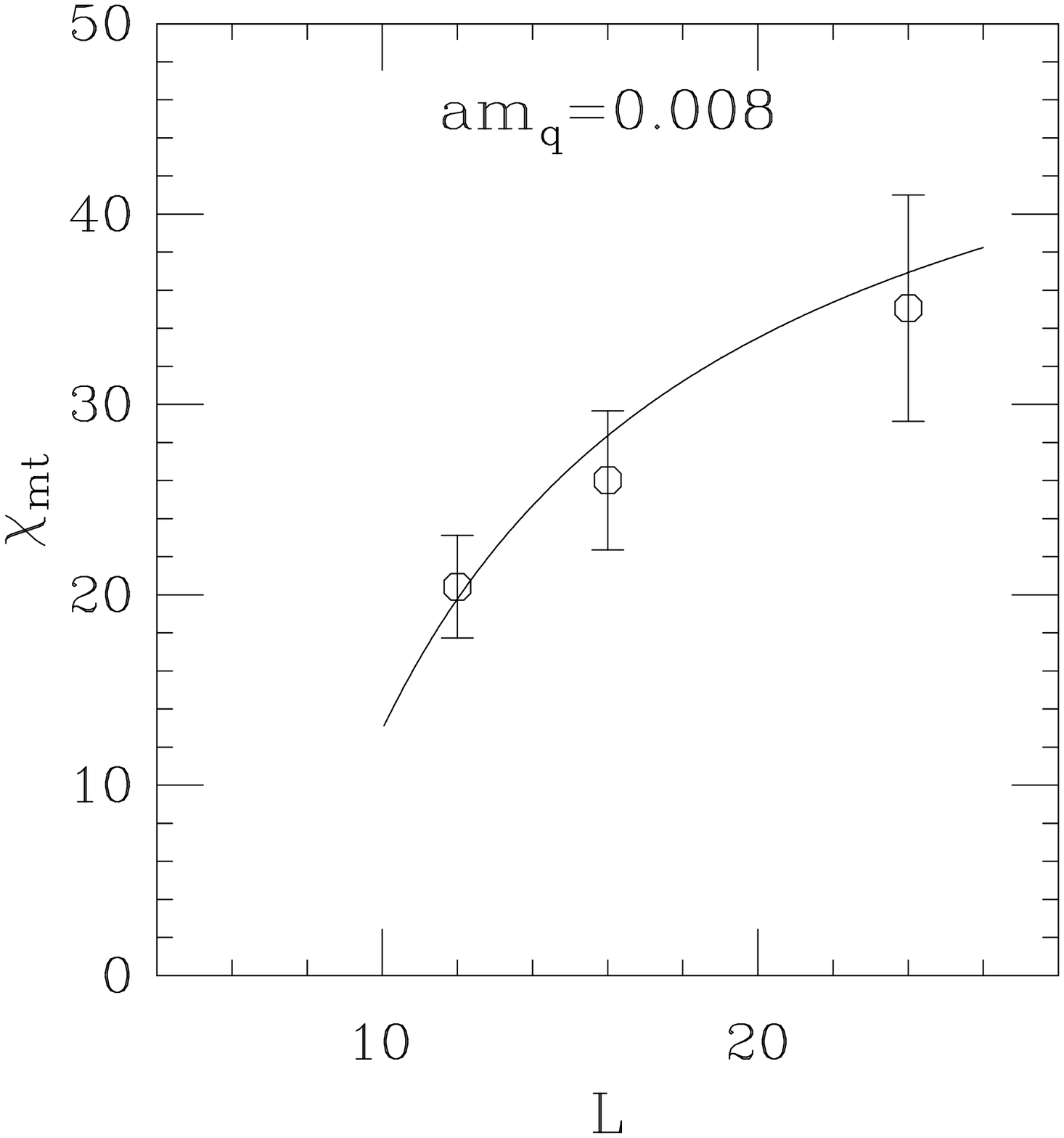,width=150mm}
\caption{Same as Fig.~\protect\ref{fig:pbpA_ext_0125}, but for $am_q =
0.008$.
\label{fig:pbpA_ext_008}
}
}
%
\figure{
 \epsfig{bbllx=100,bblly=230,bburx=530,bbury=740,clip=,
         file=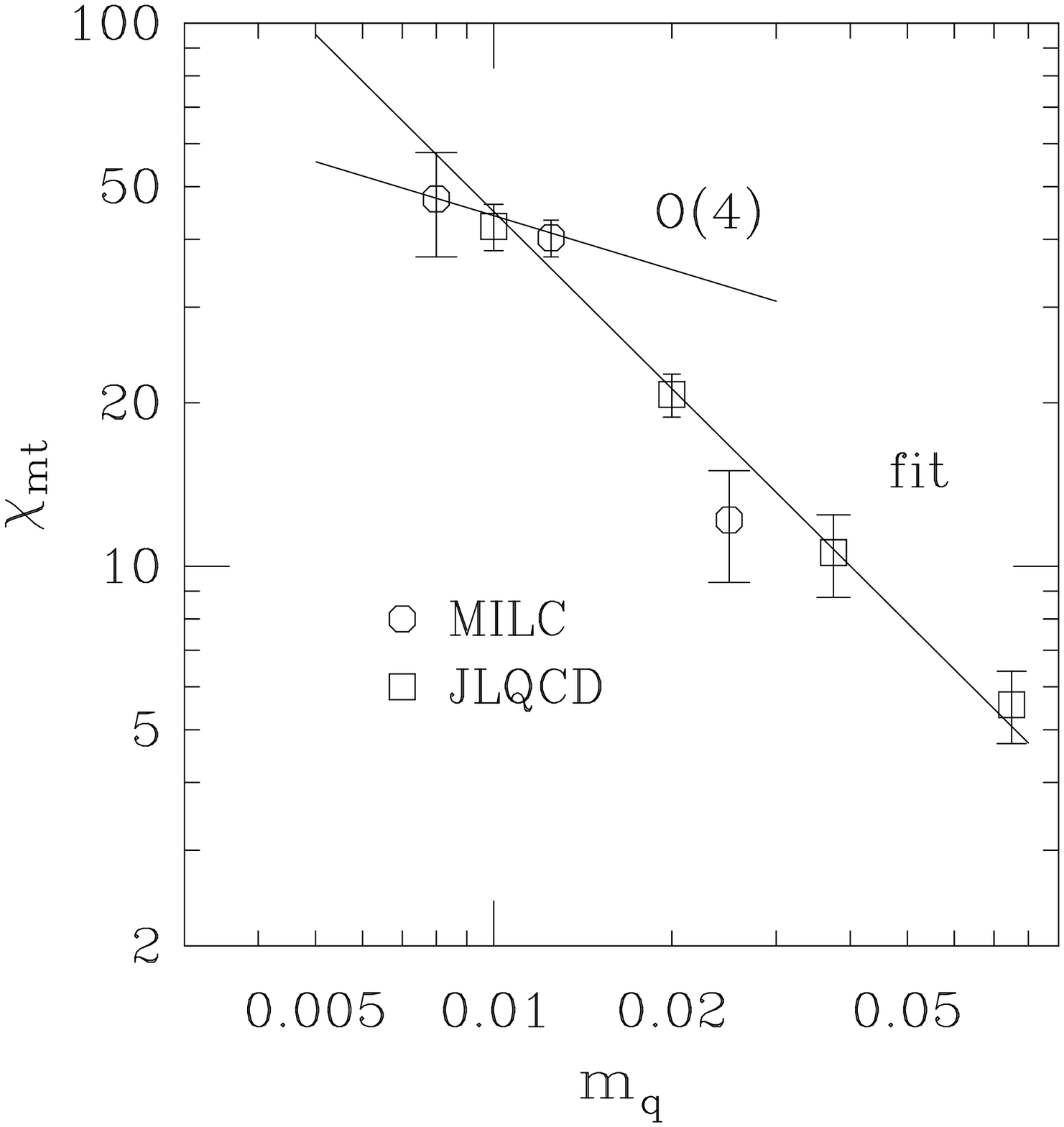,width=150mm}
\caption{Scaling analysis of extrapolated peak height for the mixed
$\pbp$--plaquette susceptibility $\chi_{mt}$ for $N_t = 4$.  Results
from Ref.~\protect\cite{JLQCD97} are obtained by similar infinite
volume extrapolation.  Also shown is the $O(4)$ scaling
prediction.
\label{fig:pbpA_ext}
}
}
%
\figure{
 \epsfig{bbllx=100,bblly=730,bburx=530,bbury=1240,clip=,
         file=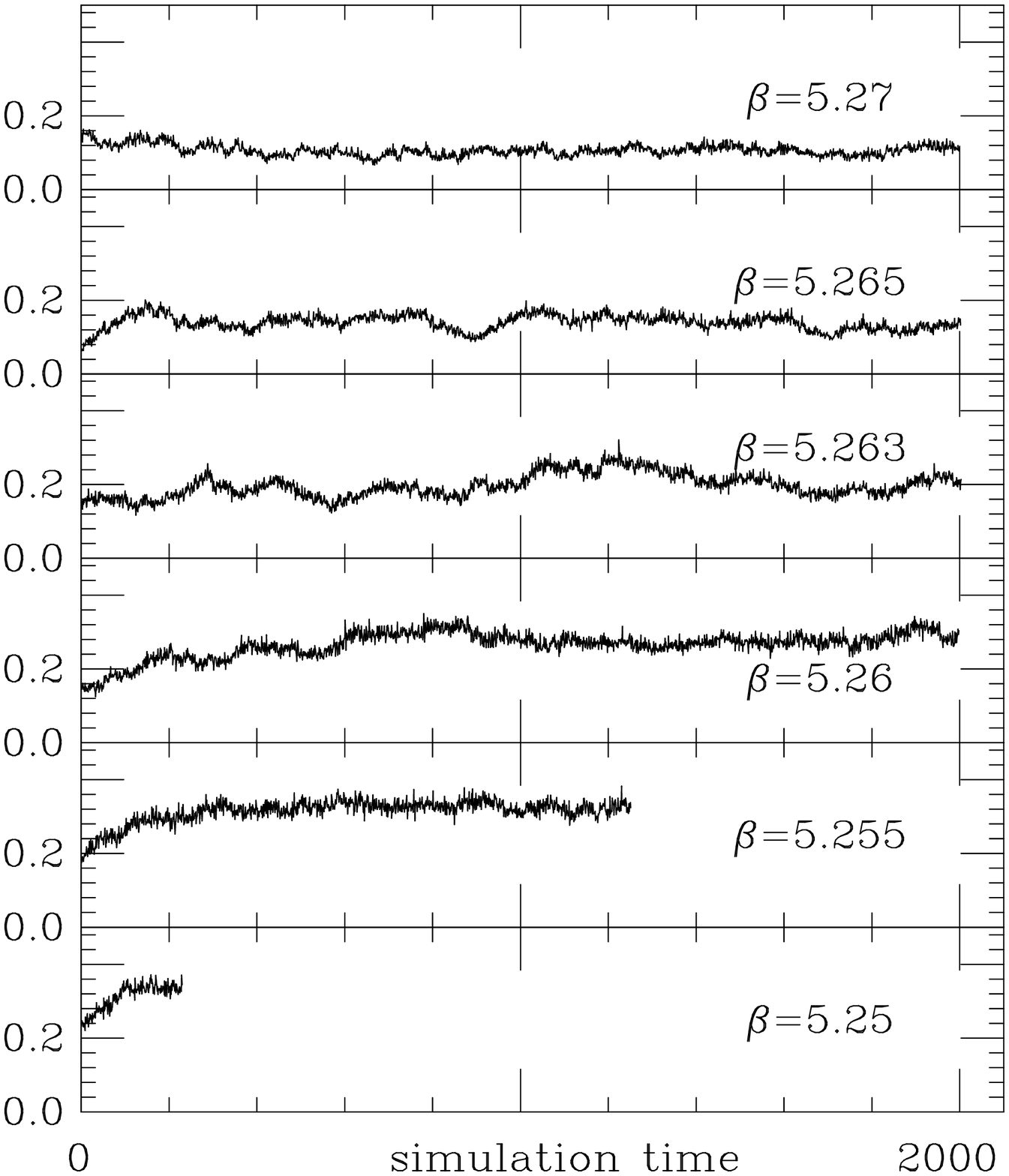,width=164mm}
\caption{Simulation time history of the order parameter $\pbp$ near
the crossover for the largest volume $24^3\times 4$.
\label{fig:pbp.m008l24}
}
}
\figure{
 \epsfig{bbllx=100,bblly=730,bburx=530,bbury=1240,clip=,
         file=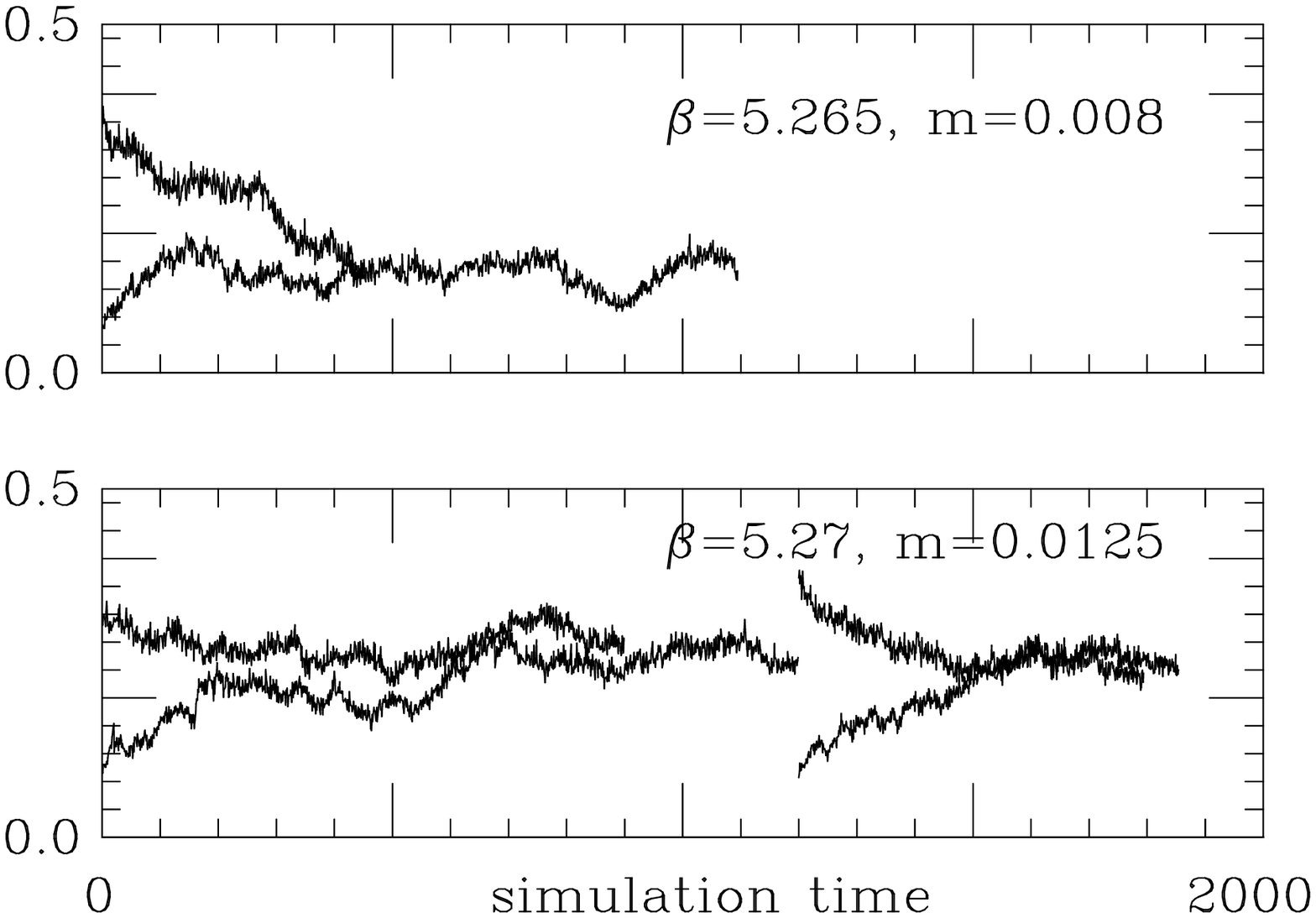,width=164mm}
\caption{Same as Fig.~\ref{fig:pbp.m008l24} showing evolution from hot
and cold starts at or near the crossover.  In each case the initially
upper (lower) trace follow a cold (hot) start.
\label{fig:pbp_hot_cold}
}
}
\figure{
 \epsfig{bbllx=100,bblly=230,bburx=530,bbury=740,clip=,
         file=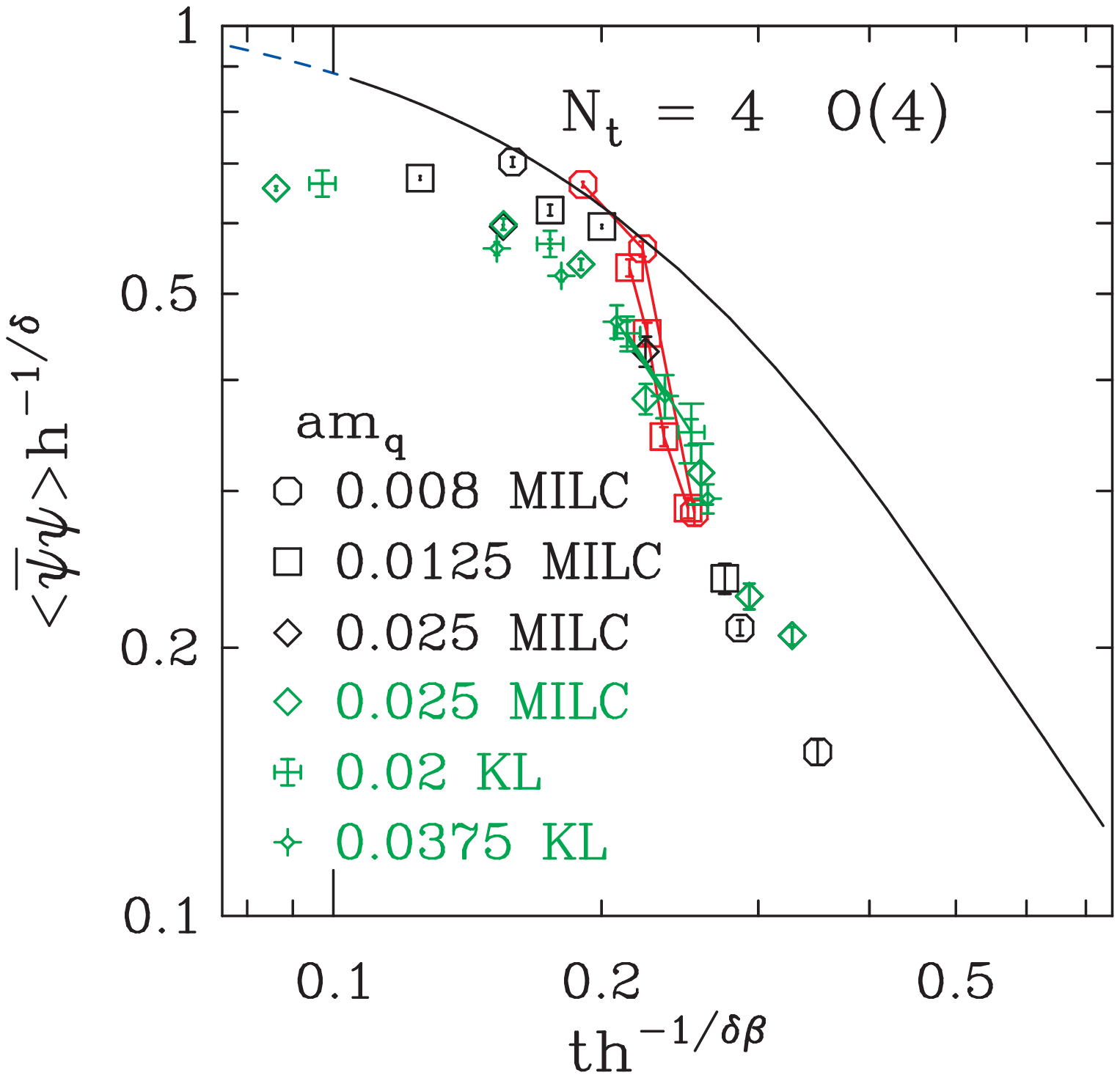,width=150mm}
\caption{Scaling test at $N_t = 4$ of the order parameter $\pbp$,
based on $O(4)$ critical exponents.  Shown for comparison is the
$O(4)$ scaling function from Ref.~\protect\cite{O4}.  The crossover region is
indicated by line segments in the data and a dashed line in the $O(4)$
scaling function. 
\label{fig:pbpnt4.change.5225}
}
}
\figure{
 \epsfig{bbllx=100,bblly=230,bburx=530,bbury=740,clip=,
         file=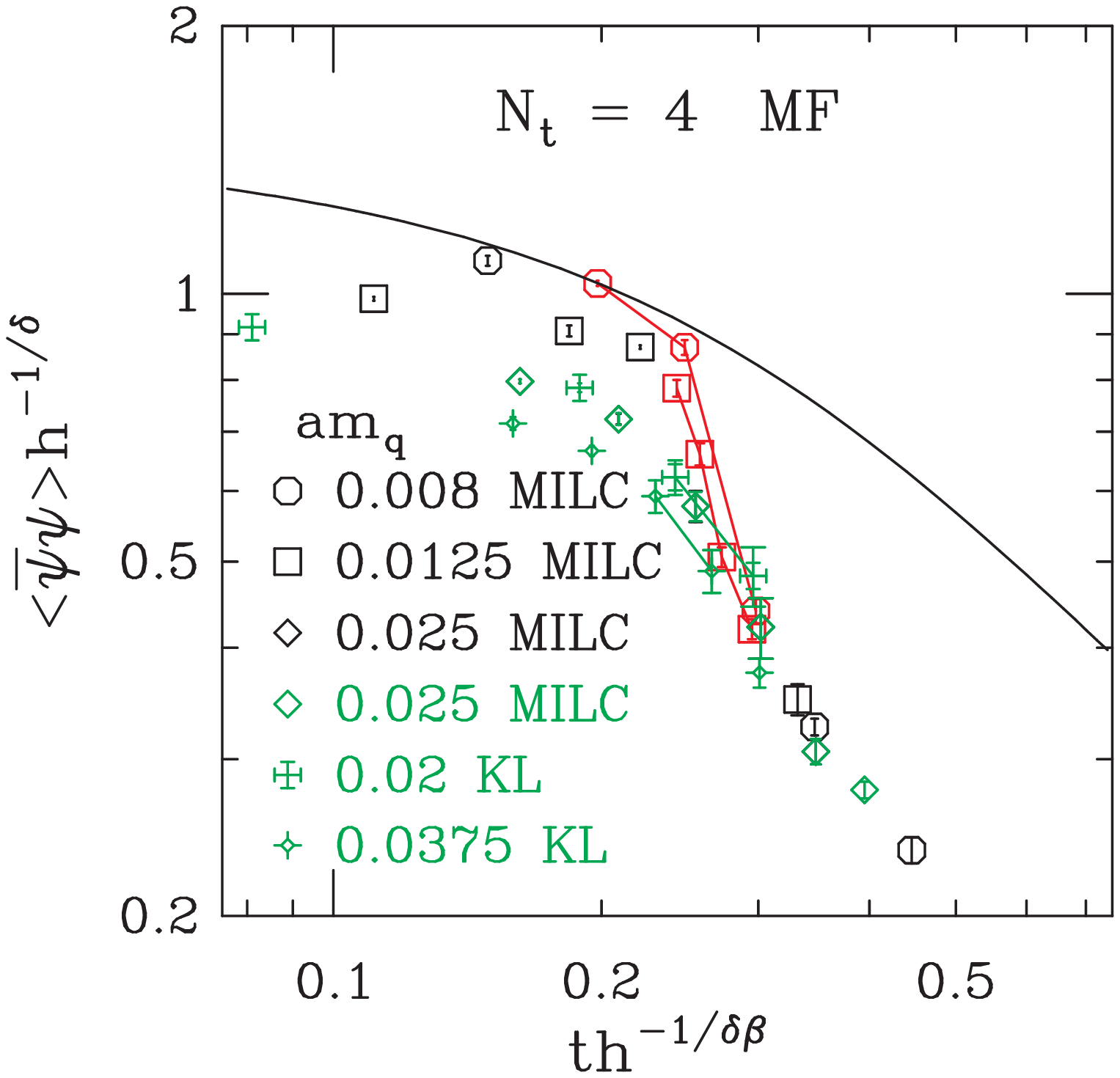,width=150mm}
\caption{Same as Fig.~\ref{fig:pbpnt4.change.5225} but with mean field
exponents and the mean field scaling function.
\label{fig:pbpnt4.change.5235.mf}
}
}
\figure{
 \epsfig{bbllx=200,bblly=130,bburx=830,bbury=940,clip=,
         file=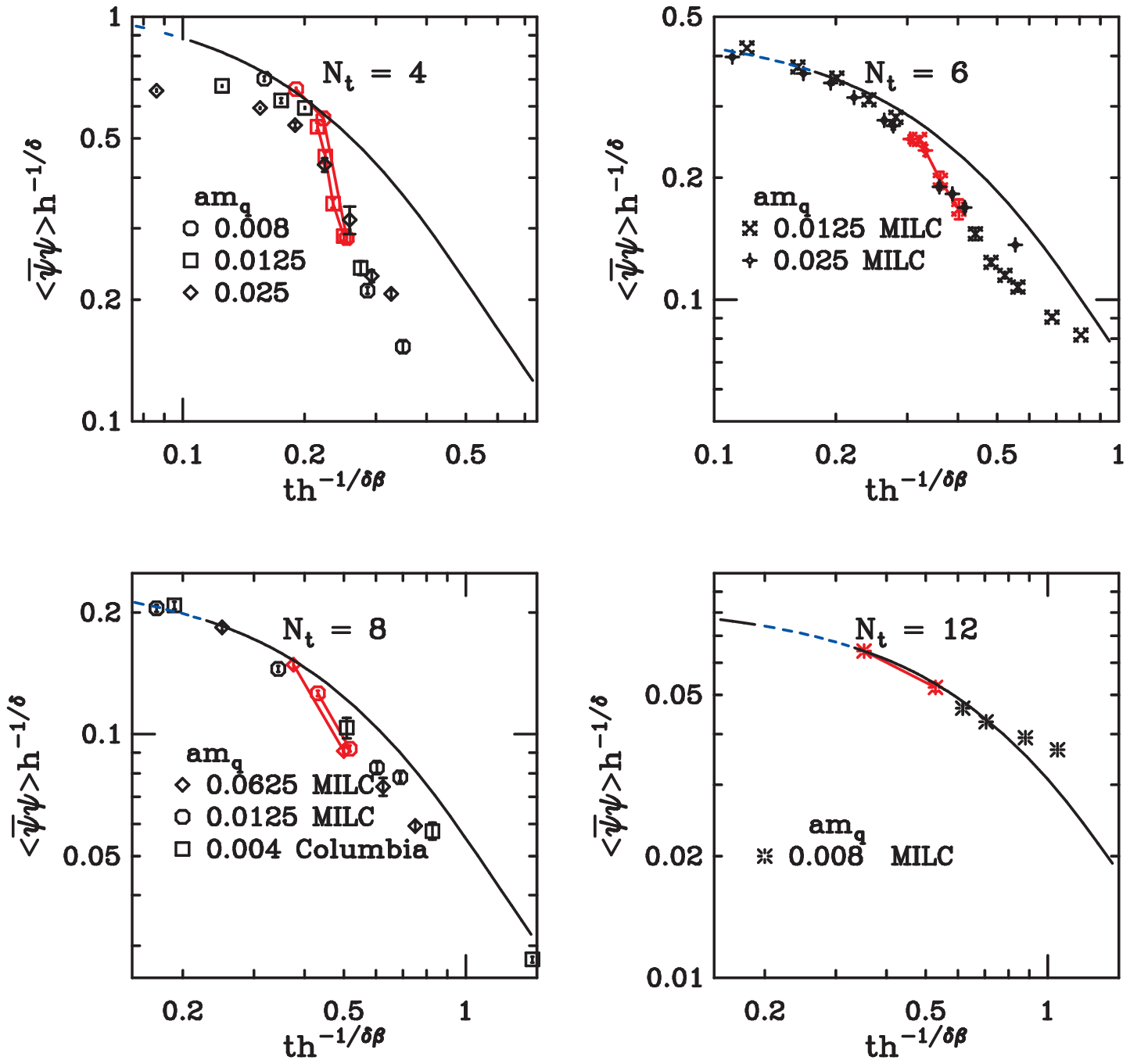,width=160mm}
\caption{Same as Fig.~\ref{fig:pbpnt4.change.5225} but for $N_t = 4$, 
6, 8, and 12.
\label{fig:pbp_o4}
}
}

\begin{references}
\bibitem{PW} R.~Pisarski and F.~Wilczek, Phys.\ Rev.\ D {\bf 29}, 338 (1984).
\bibitem{KL} F.~Karsch and E.~Laermann, Phys.\ Rev.\ D {\bf 50}, 6954 (1994).
\bibitem{Amit} D.J.~Amit, {\em Field Theory, the Renormalization Group, and
   Critical Phenomena}, (McGraw Hill, New York, 1978).
\bibitem{O4} D.~Toussaint, Phys.\ Rev.\ D {\bf 55}, 362 (1997).
\bibitem{CD97} C.~Bernard {\it et al.},
   Nucl.\ Phys.\ B (Proc.\ Suppl.) {\bf 63}, 400 (1998).
\bibitem{Nt12} C.~Bernard {\it et al}, Phys.\ Rev.\ D {\bf 54}, 4585 (1996).
\bibitem{U96} A.~Ukawa, Lattice '96, 
  Nucl.\ Phys.\ B (Proc.\ Suppl.) {\bf 53}, 95 (1997).
\bibitem{B96}
  G.~Boyd, F.~Karsch, E.~Laermann, and M.~Oevers, Talk given at 10th
  International Conference on Problems of Quantum Field Theory, Alushta,
  Ukraine, 13-17 May 1996. {\tt hep-lat/9607046} (unpublished).
\bibitem{JLQCD97} S.~Aoki {\it et al.} (JLQCD),
  Phys.\ Rev.\ D {\bf 57}, 3910 (1998).
\bibitem{KK} K.~Kanaya and S.~Kaya, Phys.\ Rev.\ D {\bf 51} 2404, (1995).
\bibitem{Wscale} Y.~Iwasaki, K.~Kanaya, S.~Kaya, T.~Yoshie, Phys.\ Rev.\
  Lett.\ {\bf 78} 179, (1997);
  S.~Aoki, Y.~Iwasaki, K.~Kanaya, S.~Kaya, A.~Ukawa, and T.~Yoshi\'e,
  Nucl.\ Phys.\ B (Proc.\ Suppl.) {\bf 63}, 397 (1998).
\bibitem{KLS4} J.~Kogut, J.-F.~Laga\"e, and D.K.~Sinclair, 
  Phys.\ Rev.\ D {\bf 58}, 034504 (1998).
\bibitem{KLS6} J.~Kogut, J.-F.~Laga\"e, and D.K.~Sinclair, 
  Nucl,\ Phys.\ B (Proc.\ Suppl.) {\bf 73}, 471 (1999).
\end{references}
\end{document}